# Prevalence of DNSSEC for hospital websites in Illinois.

Robert Robinson
Department of Internal Medicine
SIU Medicine
Springfield, IL, USA
rrobinson@siumed.edu

## Abstract

The domain name system translates human friendly web addresses to a computer readable internet protocol address.  This basic infrastructure is insecure and can be manipulated.  Deployment of technology to secure the DNS system has been slow, reaching about 20% of all web sites based in the USA.  Little is known about the efforts hospitals and health systems make to secure the domain name system for their websites.  To investigate the prevalence of implementing Domain Name System Security Extensions (DNSSEC), we analyzed the websites of the 210 public hospitals in the state of Illinois, USA.  Only one Illinois hospital website was found to have implemented DNSSEC by December, 2017.

## Introduction

The domain name system (DNS) translates a human friendly web address (www.google.com) to an internet protocol (IP) address (172.217.23.132).  The DNS infrastructure is insecure and is subject to manipulation that can result in redirection and interception of web traffic.  This vulnerability has been used for denial of service (DOS) attacks, silent redirection to another website, and man in the middle attacks (Liu, 2017; Fulton, S.  2015; Vamosi, 2011; Borgwart et al., 2015, Wang, 2015, Anagnostopoulos et al., 2013).

These attack methodologies can be uses to deny access to services, obtain login credentials or other personal information, capture information viewed over the connection, and manipulate data (Fulton, S. 2015; Geer, 2015, Borgwart et al., 2015).  These risks are particularly important for healthcare information because of the potential to delay, deny, and interfere with the delivery of medical services in addition to the potential of identity theft.

These risks can be mitigated by implementing Domain Name System Security Extensions (DNSSEC) to secure the relationship between a web address and an internet protocol address in a verifiable manner (Williamson, 2015;  Sample and Karamanian, 2015). This system allows verification that communication is occurring directly with the correct receiver.

Implementation of DNSSEC has expanded in recent years due to concerns of cybercrime and cyberwarfare.  Data published in November, 2017 showed that 24% of all websites and 90% of federal government websites implement DNSSEC (Castro, Nurko and McQuinn 2017; APNIC, 2017).  DNSSEC implementation misconfigurations are common for many popular websites (Dai, Shulman, Waidner 2016) The prevalence of DNSSEC implementation for hospitals and other healthcare entities is not known.

To investigate the implementation of this important information security measure at healthcare institutions, this study will determine the prevalence of DNSSEC implementation on the websites of public hospitals in the state of Illinois.

## Methods

This study used the DNSSEC evaluation methodology described by Castro, Nurko and McQuinn for Benchmarking U.S. Government Websites (Castro, Nurko and McQuinn 2017).

The Illinois Department of Public Health (IDPH) hospital directory was downloaded from the IDPH website [https://data.illinois.gov/dataset/410idph_hospital_directory]. This directory included all 210 licensed public hospitals in the state of Illinois as of September 27, 2017.

Each listing includes the hospital legal name, city and county the hospital is located in, and type of hospital.

IDPH defined hospital types are: general hospital, critical access hospital (no more than 25 inpatient beds), long term acute care hospital (a hospital focused on patients who have an average hospital stay of more than 25 days), pediatric hospital (exclusively serves children), psychiatric hospital (a hospital focused on psychiatric care), and rehabilitation hospital (a hospital focused on rehabilitation after stabilization of acute medical issues).

For each hospital on the list, the Google search engine (https://www.google.com) was used to identify the hospital website by searching for the hospital name, city, and state. The hospital web site uniform resource locator (URL) was entered into the DNSSEC Debugger (https://dnssec-debugger.verisignlabs.com/) developed by Verisign Labs for analysis.

DNSSEC Debugger performs a step by step validation for the entered URL. At each step, results of good, warning, or error are reported. The presence of one or more errors indicates failed DNSSEC implementation. For the purposes of this investigation, warnings were ignored.

These tests were performed on December 11, 2017.

The research protocol was reviewed by the Springfield Committee for Research Involving Human Subjects, and it was determined that this project did not fall under the purview of the IRB as research involving human subjects according to 45 CFR 46.101 and 45 CFR 46.102.

## Results

Websites could be identified for all 210 hospitals in the study sample. One hospital website was not functioning, so it could not be evaluated.

Only one hospital (0.5%) out of 209 in this sample passed DNSSEC.

## Discussion

The prevalence of DNSSEC for hospital websites in Illinois, USA is extremely low (0.5%) when compared to all websites in the United States (24%) and federal government websites (90%) using readily available testing tools (Castro, Nurko and McQuinn 2017; APNIC, 2017).

The reasons for this low prevalence of DNSSEC for hospital websites is unknown. However, not using this technology may put hospital website visitors at risk of redirection to a fraudulent website, data interception, and data manipulation using techniques successfully used against banking and other websites (Fulton, S. 2015; Geer, 2015). This potential has risks to the hospital in the form of damage to the hospital reputation and disruption of customer (patient) acquisition efforts. Mitigating these potential risks to visitors and the hospital would be desirable.

The cost/benefit ratio of DNSSEC is under debate because of costs, complexity of implementation, and the potential that improper DNSSEC implementation may prevent or delay access to websites (Lian et al., 2013; Cowperthwaite and Somayaji, 2010).

## Conclusions

DNSSEC implementation for hospital websites in Illinois is rare (1 out of 209 hospitals). This rate of DNSSEC implementation is much lower than a sample of all websites in the United States (24%) and the federal government (90%). The reasons for the low prevalence of DNSSEC for hospital websites is unknown. Further research is needed to better understand the significance of this data and the potential implications for the healthcare system.